\documentclass{PoS}

\title{High loop renormalization constants by NSPT:\\
 a status report}

\ShortTitle{High loop renormalization constants by NSPT: a status report}

\author{\speaker{Francesco Di Renzo}\\
        University of Parma and I.N.F.N.,
        Viale Usberti 7/A, I-43100 Parma, Italy\\
        E-mail: \email{direnzo@fis.unipr.it}}
\author{Luigi Scorzato\\
        ECT* and I.N.F.N.,
        Strada delle Tabarelle 286, I-38050 Villazzano (Trento), Italy\\
        E-mail: \email{scorzato@ect.it}}
\author{Christian Torrero\\
        Faculty of Physics, University of Bielefeld, D-33501 Bielefeld, Germany\\
        E-mail: \email{torrero@physik.uni-bielefeld.de}}
        
\abstract{We present an update on Numerical Stochastic Perturbation Theory projects for Lattice QCD, which are by now run on apeNEXT. As a first issue, we discuss a strategy to tackle finite size effects which can be quite sizeable in the computation of logarithmically divergent renormalization constants. Our first high loop determination of quark bilinears for Wilson fermions was limited to finite constants and finite ratios. A precise determination of $Z_P$ and $Z_S$ (and hence of $Z_m$) now becomes possible. We also give an account of computations for actions different from the standard regularization we have taken into account so far (Wilson gauge action and Wilson fermions). In particular, we present the status of computations for the Lattice QCD regularization defined by tree level Symanzik improved gauge action and Wilson fermions, which became quite popular in recent times. We also take the chance to discuss the related topic of the computation of the gluon and ghost propagators (which we undertook in collaboration with another group). This is relevant in order to better understand non-perturbative computations of propagators aiming at qualitative/quantitative understanding of confinement.}

\FullConference{The XXV International Symposium on Lattice Field Theory\\
		 July 30-4 August 2007\\
		 Regensburg, Germany}

\begin{document}

\section{Introduction}

Numerical Stochastic Perturbation Theory (NSPT) proved to be a viable tool to compute 
high loop renormalization constants (RC) for Lattice QCD \cite{NSPT_Z}. 
In the following we address a couple of the current issues in this subject.  
One is a problem: in our previous work we pointed out that 
finite volume effects can be quite important for quantities having an anomalous 
dimension (i.e. displaying log-divergencies). 
A second one is an opportunity: taking into account different actions 
(both for gluons and for fermions) does 
not imply a real overhead (there are no Feynman rules to derive). 
In the following we will report on these items, adding as an interesting aside 
a brief account on the 
gluon and ghost propagators, which, as a matter of fact, had not yet been looked 
at by NSPT.

\section{Computation of renormalization constants}

RI-MOM' is a convenient scheme to compute RC by NSPT. 
A useful example is the computation of quark bilinears. First of all 
we compute them (in Landau gauge) between external quark states at 
fixed (off-shell) momentum $p$

\[
\int dx \,\langle p | \; \overline{\psi}(x) \, \Gamma \, \psi(x) \; | p \rangle \, = \, G_{\Gamma}(pa)
\]

This step is taken in NSPT much the same way as in non-perturbative simulations. One 
then amputates ($S(pa)$ is the quark propagator) and projects out the tree-level 
structure

\[
\Gamma_{\Gamma}(pa)  =  S^{-1}(pa) \; G_{\Gamma}(pa) \; S^{-1}(pa) \;\;\;\;\;\;\;\;\;\;\;
O_{\Gamma}(pa)  =  Tr\left(\hat{P}_{O_{\Gamma}} \; \Gamma_{\Gamma}(pa)\right) 
\]

Renormalization conditions now read ($Z_q$ is the quark field renormalization constant)

\begin{equation}
	Z_{O_{\Gamma}}(\mu a, g(a)) \, \; Z_q^{-1}(\mu a, g(a)) \, \; O_{\Gamma}(pa) \Big|_{p^2 = \mu^2} \, = \, 1
\end{equation}

\section{Dealing with anomalous dimensions}

In NSPT one would like to take anomalous dimensions for granted. For RI-MOM' this is 
actually the case \cite{JG}. Let us keep this in mind while writing our master formula 
for the scalar current
\[
	( 1 - {{z_q^{(1)}}\over{\beta}} + \ldots ) \; ( 1 + {{z_s^{(1)}- \gamma_s^{(1)} \log(\hat{p}^2)}\over{\beta}} + \ldots )\; ( 1 + {{O_s^{(1)}(\hat{p}^2)}\over{\beta}} + \ldots ) \Big|_{p^2 = \mu^2} \, = \, 1	
\]
We explicitly wrote both the constant and the logarithmic contributions to renormalization constants (at first order the only log comes from $Z_S$ since one-loop quark-field anomalous dimension is zero in Landau gauge). $O_s^{(1)}(\hat{p}^2)$  is what is actually numerically measured. At one-loop order we can solve the previous relation to

\begin{equation}\label{Zs}
	z_q^{(1)} - z_s^{(1)} =   O_s^{(1)}(\hat{p}^2) - \gamma_s^{(1)} \log(\hat{p}^2) \,.
\end{equation}

The quantity on l.h.s. is now finite and the only dependence on $pa\equiv \hat{p}$ is an 
irrelevant one, which can be wiped out by extrapolating to zero by means of what we call 
an Hypercubic-symmetric Taylor expansion. Unfortunately this program results in a failure: see Fig.~(1) (left).

\begin{figure}[htb]
  \begin{center} 
		\includegraphics[scale=0.5]{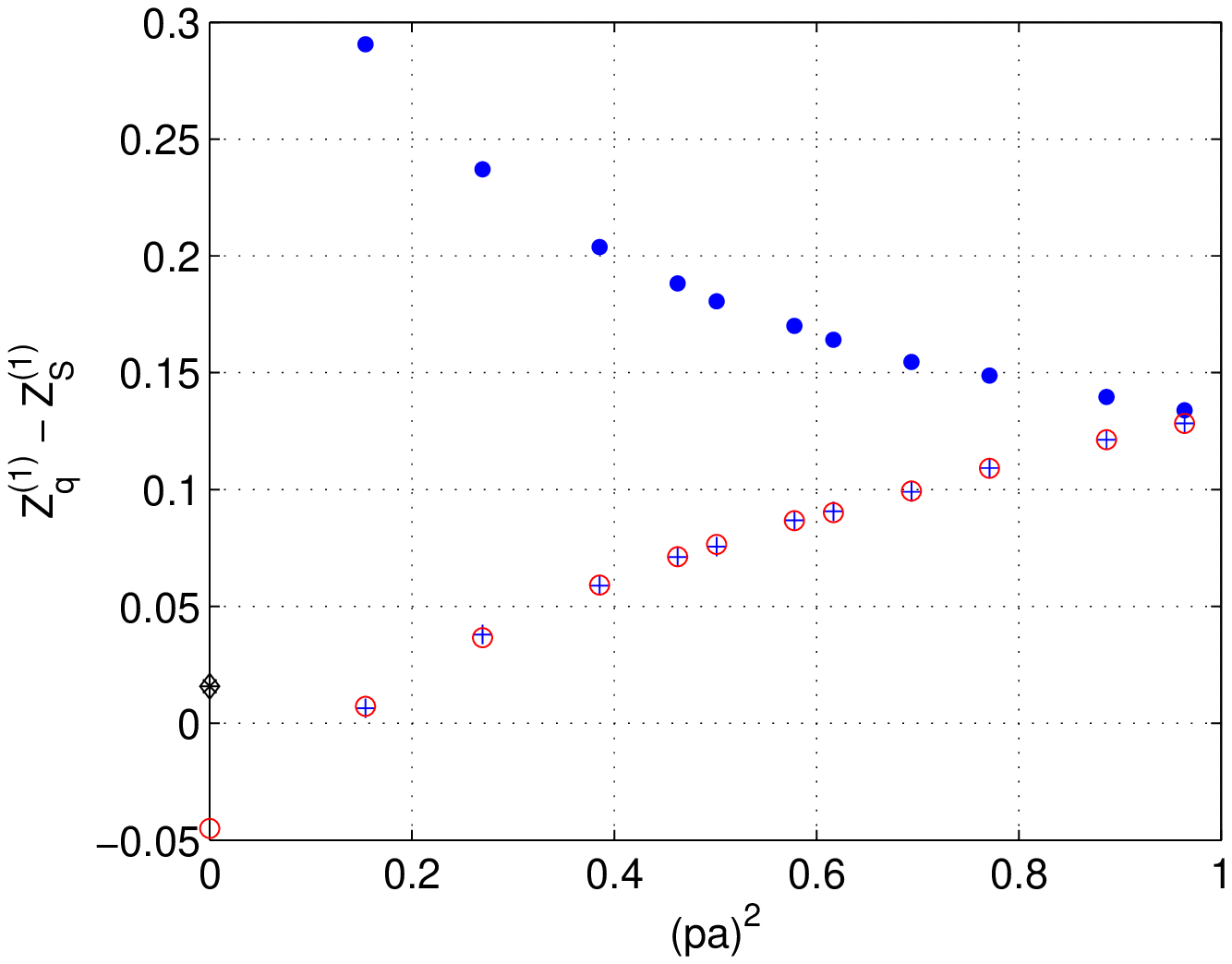}
		\includegraphics[scale=0.5]{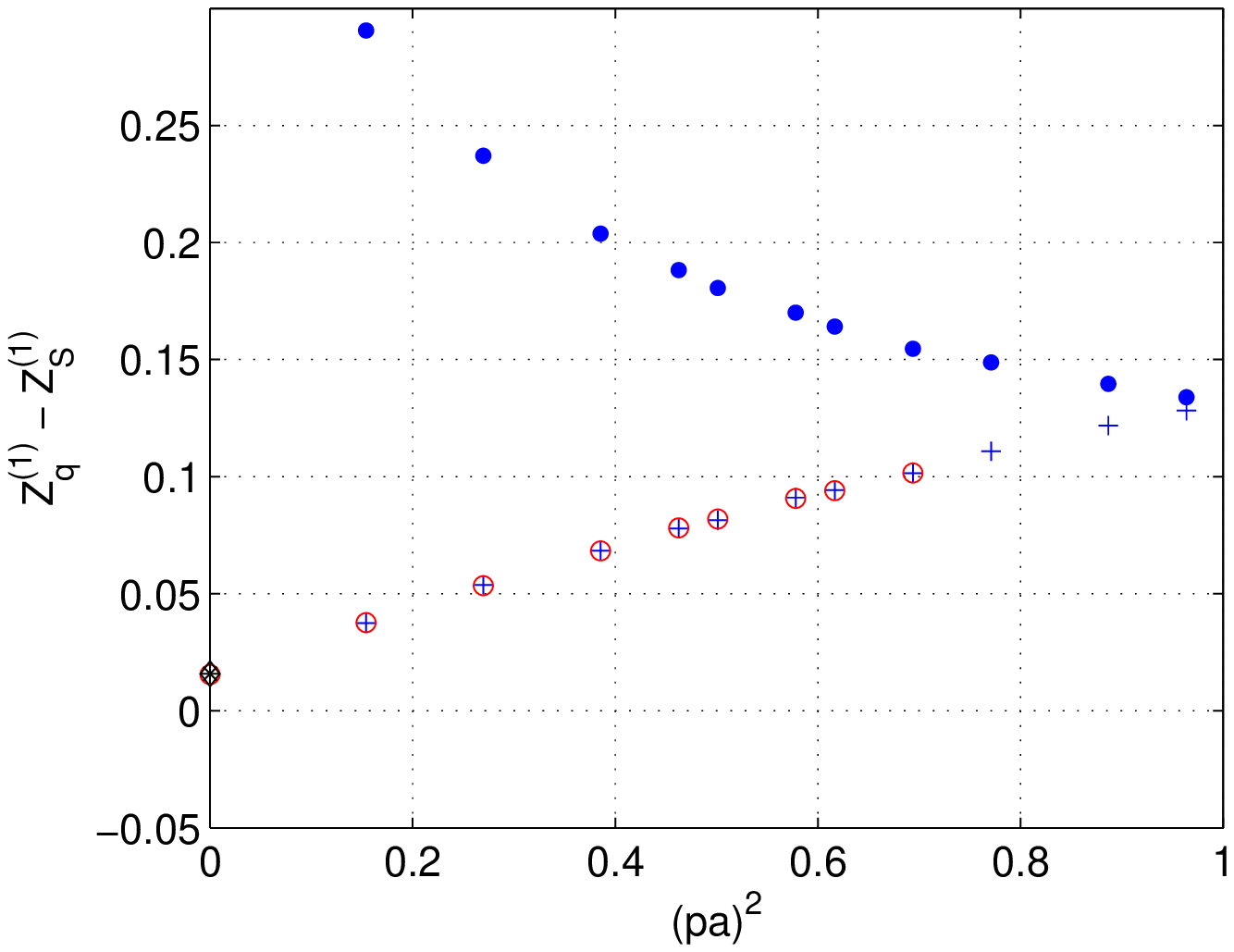}
\caption{
On the left: computation of one loop renormalization constant for the scalar current. In the 
notation of Eq.~(3.1), 
upper points are the unsubtracted $O_s^{(1)}(\hat{p}^2)$, while lower (circled  
crosses) stand for the subtracted $O_s^{(1)}(\hat{p}^2) - \gamma_s^{(1)} \log(\hat{p}^2)$.  Analytic result (darker symbol) is missed.
On the right: the same computation subtracting the \emph{tamed-log} contribution referred to in the text. Analytic result is got.}
   \label{fig1}
  \end{center}
\end{figure}

IR problems do not come as a surprise, due to finite size effects. In Fig.~(2) we display results on sizes $32^4$ and $16^4$. Results perfectly agree for the ratio of $Z_s$ and $Z_p$ (which is a finite quantity). On the other hand, the smaller the lattice, the bigger the IR effects one gets for log-divergent quantities. These are one loop results, but the picture stays much the same at higher loops.

\begin{figure}[hb]
\includegraphics[scale=0.65]{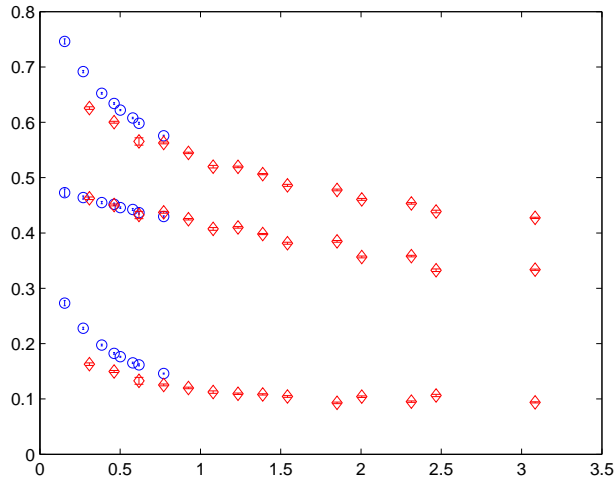}
\centering
\caption{
Computations of $O_p^{(1)}(\hat{p}^2)$ (the equivalent of Eq.~(3.1) for the pseudoscalar current) (top) and $O_s^{(1)}(\hat{p}^2)$ (bottom) on $32^4$ (circles) and $16^4$ (diamonds). In the middle the ratio $\frac{O_s^{(1)}(\hat{p}^2)}{O_p^{(1)}(\hat{p}^2)}$.}
\label{fig2} 
\end{figure}

Circumventing the problem is not so hard at one loop. Define $L = N a$. Let us now write down for the quantity at hand the momentum sum $I(p,a,L)$ of conventional Lattice Perturbation Theory,  making use of the same regularization of zero modes that we use in NSPT 
(zero momentum removed from the sum). In the same spirit of \cite{Kawai} let us now split it as
\begin{equation}
	I(p,a,L) = I(0,a,L) + \left( I(p,a,L)- I(0,a,L) \right) \equiv I(0,a,L) + J(p,a,L).
\end{equation}
The divergence is logarithmic so that by subtracting $I(0,a,L)$  we make 
$J(p,a,L)$ UV finite. IR divergences will pop up and cancel with those in $I(0,a,L)$:
\begin{eqnarray}
	I(0,a,L) & = & c_1 + \gamma \, log(a/L) + H(a/L) \\ \nonumber
	J(p,a,L) & = & c_2 + \gamma \, log(pL) + G(pa,a/L,pL)\,.
\end{eqnarray}
Now we look for $pL=\hat{p}N$ effects.
These should be looked for in $G(pa,a/L,pL) \rightarrow \tilde{G}(pL)$, which we computed from the formal continuum limit of our sum  $J(p,a,L)$: this means $a \rightarrow 0$  with $L = N a$  fixed 
(as we pointed out, $J(p,a,L)$ is UV finite). Again, we need the same \emph{ad hoc} regularization of zero modes. 
We call this contribution a \emph{tamed-log}. 
It is supposed to resemble the expected log, but with $pL = \hat{p}N$ effects on top of it: 
it indeed approaches a log for $p>>1$. By subtracting this \emph{tamed-log} we got the right one loop result.

\section{Dealing with anomalous dimensions: a new strategy.}
\vspace{0.4cm}

The above method has got obvious drawbacks. First of all, 
one has to go back to diagramatic computations we would have liked to get rid of 
by means of NSPT.
While one is happy enough with the one loop picture, at higher loops the situation is less clear and it is for sure much more cumbersome.
Moreover, one is not actually making use of the computations on different lattice sizes, but has to revert instead to a continuum computation.
In the following we present a strategy which is currently under investigation 
(we do not yet display results). 
This method is a very close relative of the one which is employed in an NSPT 3-d 
application with a mass in play \cite{Chr}. 

It can be better understood having in mind figure 2. 
An obvious way of looking at it is the following: on $32^4$ and $16^4$ one gets results 
at the same physical momentum $p$ from points affected by different $pL=\hat{p}N$ effects. 
Moreover, for $pL$ large enough one can get results substatially 
free of finite volume effects.
Having measurements on at least two different sizes ($N_1$, $N_2$, $N_1>N_2$), this suggests to proceed as follows:
\begin{itemize}
\item
Go to a momentum $p$ high enough: the measurement $f(\hat{p},\hat{p} N_1 >>1) 
\approx f(\hat{p},\infty)$ on the big lattice will be substantially 
free of finite size effects.
\item 
Read the deviation $\delta(\hat{p} N_2) = f(\hat{p},\hat{p} N_2) - f(\hat{p},\infty)
\approx f(\hat{p},\hat{p} N_2) - f(\hat{p},\hat{p} N_1)$
(\emph{i.e.} the deviation from the result you got on the little lattice at the same value of $p$).
\item
This deviation defines the $pL=\hat{p}N$ effect you have to correct for 
at the point having the same value of $pL=\hat{p}N$.
\end{itemize}
The strategy is promising and first results are encouraging.

\section{NSPT computations for different actions}

Changing the action can be quite cumbersome for Lattice Perturbation Theory: deriving 
Feynman rules for propagators and (both relevant and irrelevant) vertices can require hard work (this is usually a computer-aided task). Life is by far easier in NSPT. We are working on different Lattice QCD regularizations resulting from taking different gauge and fermionic actions. For the gauge 
action we consider plain Wilson, Iwasaki or (tree-level) Symanzik. For fermionic action 
we take plain Wilson or Clover. From the computational point of view, notice that all the combinations are fairly well implemented on apeNEXT (as compared to APEmille) because 
of the larger amount of available memory.

\begin{figure}[t!]
\includegraphics[scale=0.53]{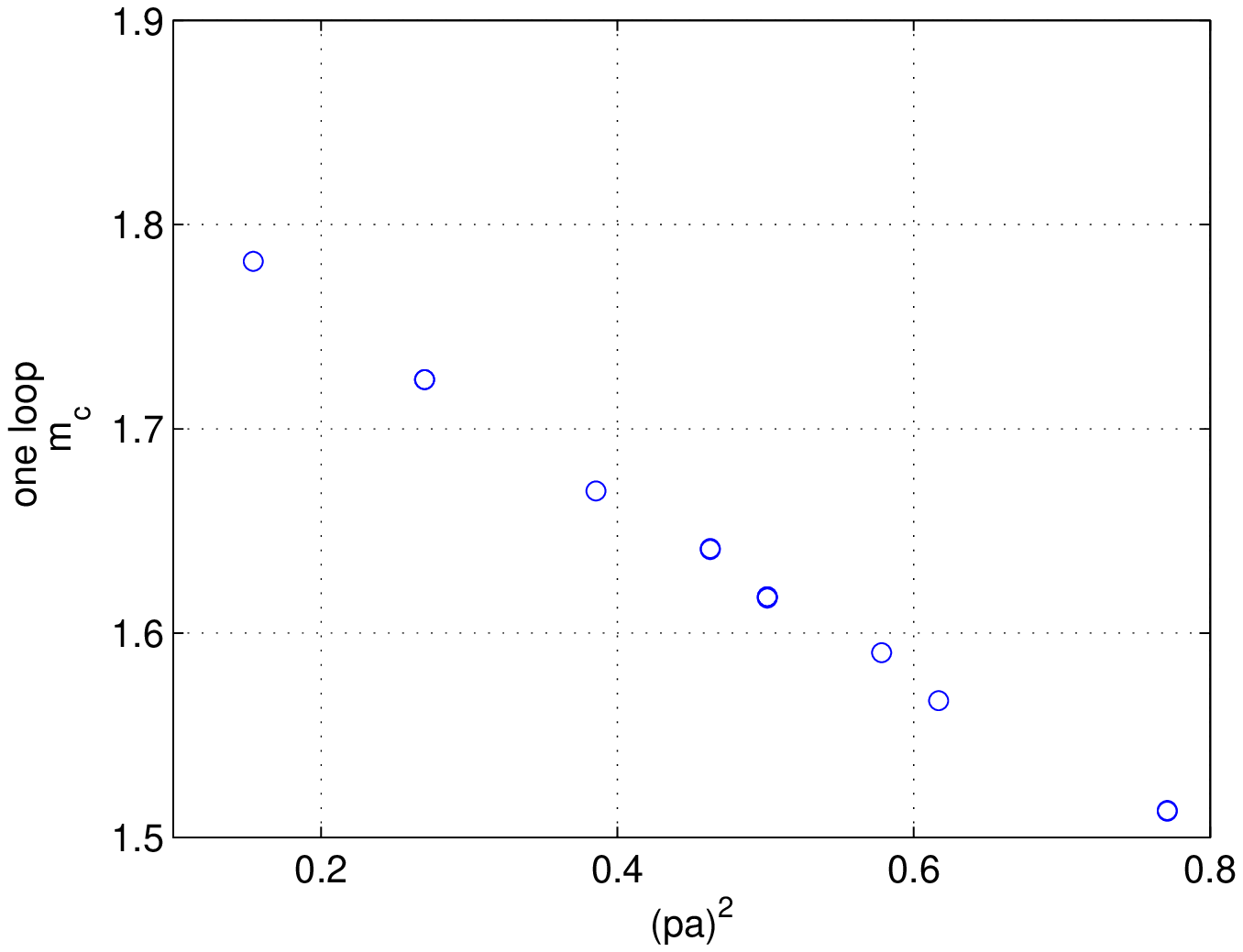}
\includegraphics[scale=0.53]{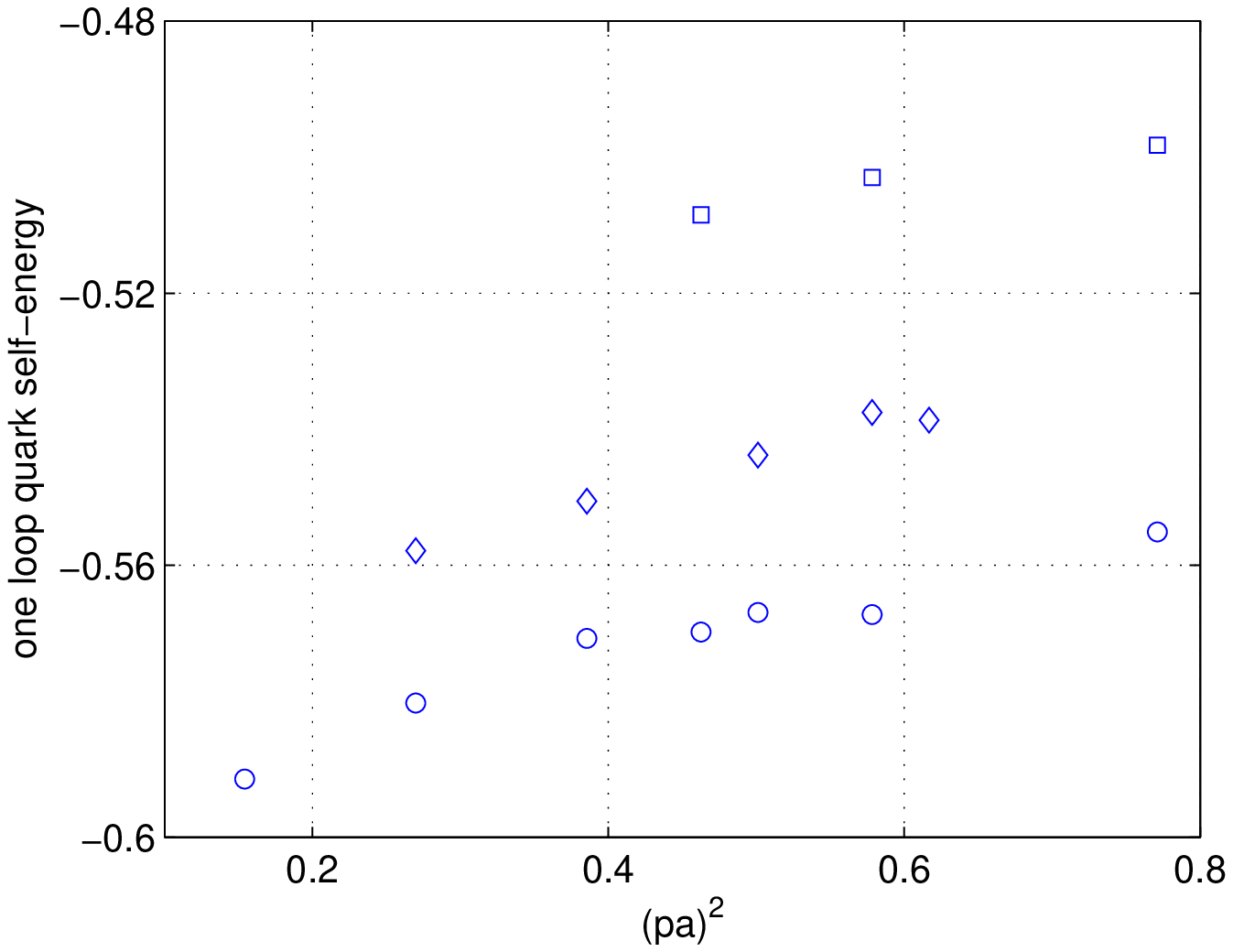}
\caption{
Computations of the Wilson fermions one loop critical mass (left) and self-energy (right) for tree-level Symanzik improved gauge action. For the self-energy points come in "families" corresponding to different violations of Lorents symmetry, depending on the lenght of momentum along the gamma matrix one traces with. Results from the measurement on a single configuration.
See text for an important caveat.}
\label{figSym} 
\end{figure}

Fig.~(3) displays one loop benchmark computations of critical mass and fermion self-energy for the tree-level Symanzik improved gauge action with plain Wilson fermions.
At the time of the conference we were still in a bit embarassing situation: from Fig.~(3) one can inspect results were (slightly) off the analytic results available  
in the literature \cite{Aoki,Haris} (a wrong normalization is now being fixed). Notice with this respect that a one loop benchmark is a very safe one for NSPT: since everything is implemented order by order, once you get first order right you are standing on a quite safe ground.
Apart from the actual results, notice that plots are measurements from a single configuration.
Fermionic measurements are remarkably stable (curves are smooth enough): keep this in mind now that we move to the Gluon Propagator.

\section{Gluon (and Ghost) Propagator}

Gluon and ghost propagators are the subject of a new collaboration\footnote{
F.~Di Renzo, C.~Torrero, M.~Ilgenfritz, M.~Muller-Preussker, H.~Perlt, A.~Schiller.}. 
The framework for this work is the qualitative/quantitative investigation of 
confinement via Schwinger-Dyson equations for the fundamental degrees of freedom of the theory.\\
The gluon propagator in NSPT is a really straightforward computation. 
One first fixes the gauge: once again, Landau is the obvious choice 
(FFT acceleration is a powerful tool at hand).
One then goes to momentum space, takes the product of fields and traces.
The ghost propagator is measured much the same way as 
the fermion propagator, by inverting order by order the Faddeev-Popov matrix on a (momentum space) source. 
The technology for the ghost was derived in the context of another application \cite{MikkoYork}. 

\begin{figure}[t!]
\includegraphics[scale=0.53]{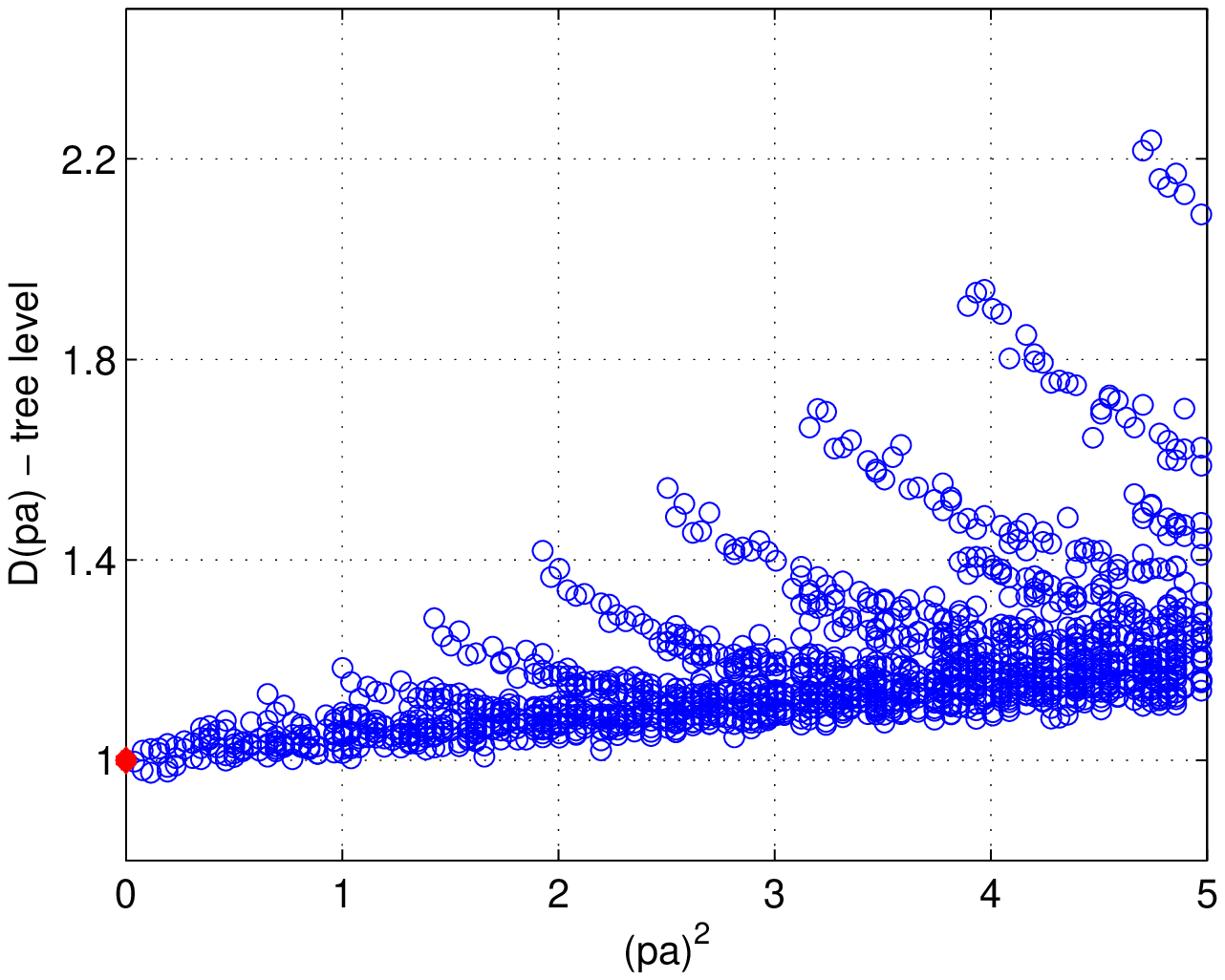}
\includegraphics[scale=0.53]{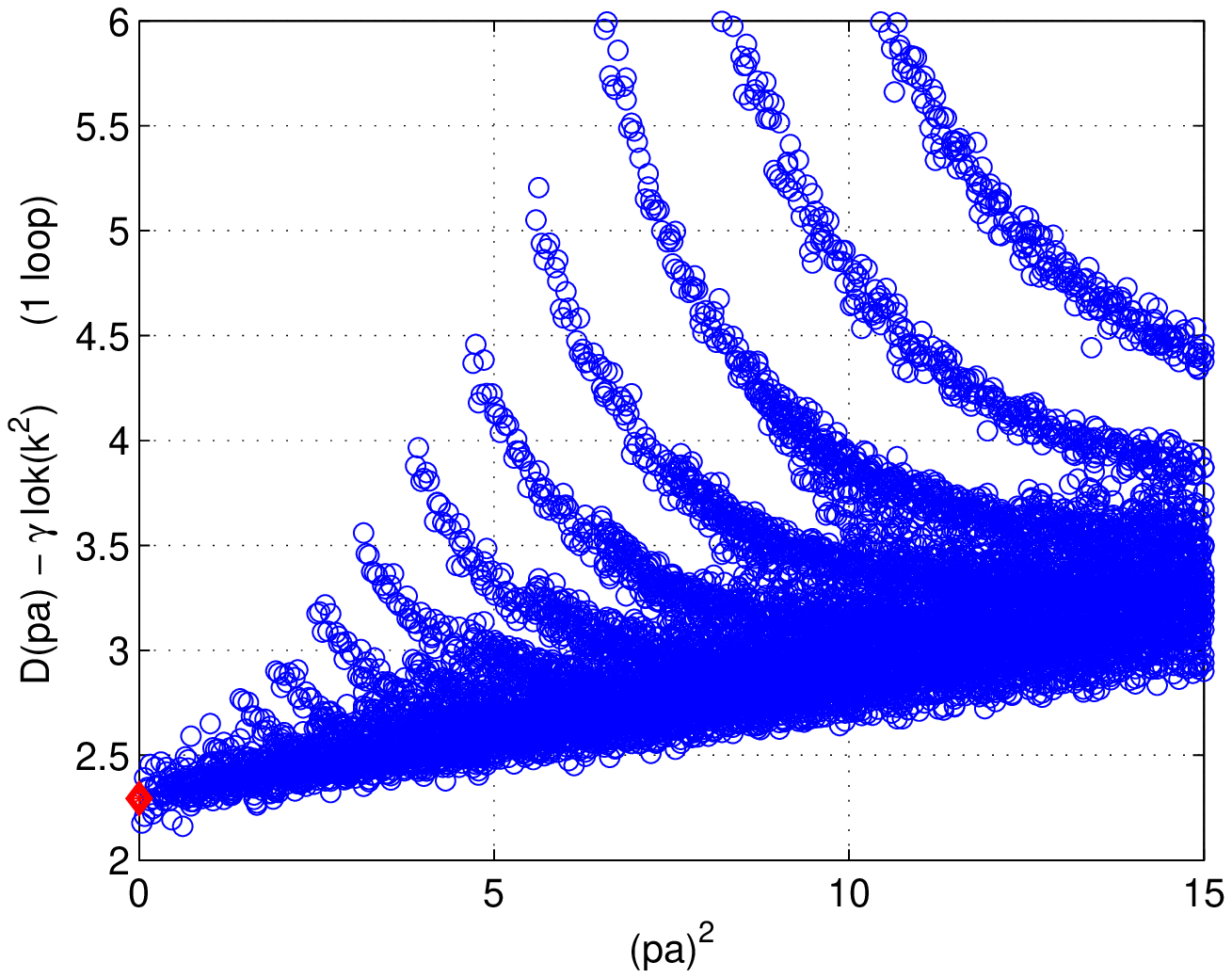}
\caption{
Tree level (left) and one loop (right) computation of the gluon propagator. Notation is: $\Pi_{\mu\nu}(k)=D(k)\,\frac{\left(\delta_{\mu\nu}-\frac{k_\mu\,k_\nu}{k^2}\right)}{k^2}$. 
The analytically known constants are recovered in 
the continuum limit (at one loop the log dictated by anomalous 
dimension has to be subtracted). Also in this case data arrange themselves in families according 
to different patterns of violation of Lorentz symmetry.}
\label{figProp} 
\end{figure}

In figure 4 we display tree-level and one loop benchmark computations of the gluon propagator. 
Notice that this time we present also a tree level computation as a benchmark. The gluon propagator is with this respect quite different from the fermion (or ghost) propagator:
there is no inversion on a source, but simply the computation of a 
correlator (in momentum space). While this is straightforward, one now gets much 
more noise. Even tree level comes from an actual measurement: first fluctuations 
around the vacuum are the Lie algebraic free 
fields whose correlatator gives the free Feynman propagator.
It is funny to compare the previous plots (which come out of $O(100)$ configurations) with 
the (one configuration) measurements of Fig.~(3).
Still, although noisy, this is a doable computation. Both for the gluon and for the ghost, we 
provide measurements from fairly big lattices ($16^4$ and $32^4$). These measurements 
are taken on the same configurations stored to measure the QCD fermionic quantities. 
Our German collaborators \cite{Misha} provide instead huge statistics measurements from smaller lattices.

\section{Conclusions}

We have presented a new method to correct for finite volume effects in NSPT renormalization 
constants computations. The method is much less cumbersome than the one 
we applied to one loop computations. \\

We are computing renormalization constants for different regularizations of Lattice QCD: computations for Iwasaki and tree level Symanzik improved gauge actions are on their way. \\

We also reported on the computation of the gluon (and ghost) propagator. While 
the gluon propagator is the most straightforward to measure, it is not at all the best one 
with respect to the statistical fluctuations.


\begin{thebibliography}{99}
\bibitem{NSPT_Z}
  F.~Di Renzo, V.~Miccio, L.~Scorzato and C.~Torrero,
  \emph{High-loop perturbative renormalization constants for Lattice QCD (I):
  finite constants for Wilson quark currents},
  \emph{Eur.\ Phys.\ J.} {\bf C 51} (2007) 645 
  [{\tt hep-lat/0611013}].
\bibitem{JG}
  J.A. Gracey,
  \emph {Three loop anomalous dimension of non-singlet quark currents in the RI' scheme},
  \emph{Nucl.\ Phys.} {\bf B 662} (2003) 247
  [{\tt hep-ph/0304113}].
\bibitem{Kawai}
  H.~Kawai, R.~Nakayama and K.~Seo,
  \emph{Comparison of the Lattice Lambda parameter with the continuum Lambda
  parameter in massless QCD}, 
  \emph{Nucl.\ Phys.}  {\bf B 189} (1981) 40.
\bibitem{Chr}
  C.~Torrero, M.~Laine, Y.~Schroeder, F.~Di Renzo, V.~Miccio, 
  \emph{Towards 4-loop NSPT result for a 3-dimensional
	condensate-contribution to hot QCD pressure}, these proceedings.
\bibitem{Aoki}
	S.~Aoki, K.~Nagai, Y.~Taniguchi, A.~Ukawa, 
	\emph{Perturbative renormalization factors of bilinear quark operators for
  improved gluon and quark actions in lattice QCD},
	\emph{Phys. Rev.} {\bf D 58} (1998) 074505 
	[{\tt hep-lat/9802034}].
\bibitem{Haris}
  M.~Constantinou, H.~Panagopoulos and A.~Skouroupathis,
  \emph{Improved perturbation theory for improved lattice actions},
  \emph{Phys.\ Rev.} {\bf D 74} (2006) 074503 
  [{\tt hep-lat/0606001}].
\bibitem{MikkoYork}
  F.~Di Renzo, M.~Laine, V.~Miccio, Y.~Schroder and C.~Torrero,
  \emph{The leading non-perturbative coefficient in the weak-coupling expansion  of
  hot QCD pressure},
  \emph{JHEP} {\bf 0607} (2006) 026 
  [{\tt hep-ph/0605042}].
\bibitem{Misha}
  M.~Ilgenfritz, H.~Perlt, A.~Schiller, 
  \emph{The lattice gluon propagator in stochastic perturbation theory}, these proceedings.
\end{thebibliography}
\end{document}